\documentclass[prl,twocolumn]{revtex4}%
\usepackage{amsfonts}
\usepackage{amsmath}
\usepackage{amssymb}
\usepackage{graphicx}%
\providecommand{\U}[1]{\protect\rule{.1in}{.1in}}

\begin{document}
\preprint{ }
\title[Short title for running header]{Giant magnetic broadening of ferromagnetic resonance in a GMR Co/Ag/Co/Gd quadlayer.}
\author{S. Demirtas}
\affiliation{University of Texas at Dallas}
\author{A. R. Koymen}
\affiliation{University of Texas at Arlington}
\author{M. B. Salamon}
\affiliation{University of Texas at Dallas}
\keywords{ferromagnetic resonance, giant magnetoresistance, torque-correlation}
\begin{abstract}
Both magnetic-resonance damping and the giant magnetoresistance effect have
been predicted to be strongly affected by the local density of states in thin
ferromagnetic films. We employ the antiferromagnetic coupling between Co and
Gd to provide a spontaneous change from parallel to antiparallel alignment of
two Co films. \ A sharp increase in magnetic damping accompanies the change
from parallel to antiparallel alignment, analogous to resistivity changes in
giant magnetoresistance.

\end{abstract}
\date{\today}

\pacs{75.40.Gb, 75.47.De, 75.50Gg,75.76.+j, 76.50.+g}
\maketitle

The discovery of giant magnetoresistance (GMR) by Baibich et al.\cite{baibich}
has led to important applications in magnetic recording and data storage.
\ Nonetheless, a fundamental understanding of the microscopic mechanism
remains a subject of continuing research.\cite{zahn,binder01} \ Early
work\cite{camley,levy} considered spin-dependent scattering to be the primary
mechanism for GMR\ effects, and indeed such scattering can considerably
enhance them.\cite{zahn95} \ However, Schep, et al.\cite{schep} were the first
to demonstrate that significant GMR (for currents perpendicular to the
magnetic layers (CPP) at least) is possible in a perfect magnetic
superlattice, a consequence of s-d hybridization and resultant differential
localization of electronic states betweeen parallel (P) and antiparallel (AP)
alignment. The same quantum-well states strongly modify the effectiveness of
scatterers at the interface\cite{binder01}, thereby contributing to GMR for
in-plane currents (CIP) as well. \ The aim of this paper is to provide
independent evidence for substantial changes in the local density of states
accompanying a transition from P to AP alignment. Exploiting the strong
antiferromagnetic coupling between Co and Gd, we fabricated a GMR structure
that spontaneously reverses the relative orientation of two Co layers as the
temperature is reduced. Upon reversal from P to AP alignment, the width of the
ferromagnetic resonance line of the free Co layer sharply changes its
temperature dependence. We interpret these results in the context of the
so-called torque-correlation model of ferromagnetic damping, \cite{kunes,
gilmore07,gilmore08}\ applicable to Co, in which the linewidth is directly
related to the local density of states; by analogy, we term the increased
broadening Giant Magneto-Broadening (GMB).

We have prepared a trilayer structure of Co/Ag/Co with an underlying Gd layer;
the Ag layer is sufficiently thick that there is no exchange coupling of the
two Co layers. \ Co and Gd are strongly coupled
antiferromagnetically.\cite{demirtas05} \ Above, and somewhat below, the Curie
temperature of Gd, the two Co layers are ferromagnetically aligned in a modest
magnetic field. \ As the temperature is reduced, the magnetic moment of Gd
increases. Below the compensation temperature $T_{comp}$, the Gd moment
exceeds that of its adjacent Co layer, causing it to align with the magnetic
field, producing AP\ alignment of the two Co layers. \ In Fig. 1, we show the
low-field magnetization of a Ag(10 nm)/Co(4nm) bilayer on Gd(10 nm). \ The
minimum in net magnetization at $T_{comp}=$ 170 K reflects the point at which
the magnetization of the underlying Gd and its adjacent Co layer are equal and
opposite, oriented perpendicular to the applied field. \ The small
paramagnetic moment at $T_{comp}$ results from the canting of the opposing
moments toward the applied field direction. \
\begin{figure}
[ptb]
\begin{center}
\includegraphics[
trim=0.000000in 0.000000in 0.019050in -0.182524in,
height=2.9386in,
width=3.5674in
]%
{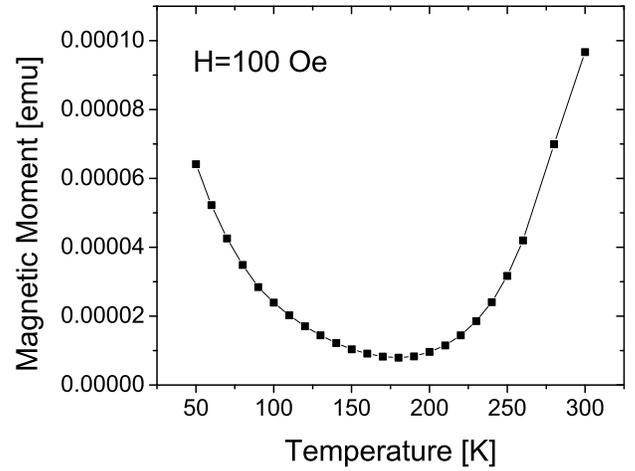}%
\caption{ Magnetic moment as a function of temperature for the [Co 40 \AA /Gd
100 \AA ] bilayer. Minimum corresponds to $T_{comp}$.}%
\end{center}
\end{figure}

Multilayer samples were created at room temperature using a dc magnetron
sputtering system. The base pressure of the deposition chamber was $10^{-9}$
Torr. Ultra high purity argon gas was used and the deposition pressure was 3
mTorr. An \textit{in situ} quartz thickness monitor, calibrated by a stylus
profilometer, measures the deposition thicknesses. Samples were sputtered from
pure Gd, Co and Ag targets on Si (100) substrates. Ag layers 200 Angstrom
(\AA ) thick were used as buffer layers in all samples. The Co(1)/Ag/Co(2)/Gd
multilayer was created with a 100 \AA \ nonmagnetic Ag spacer between the two
4 nm-Co layers, thick enough to suppress any long range exchange interactions.
\ A 100-\AA \ Ag cap layer completed the deposition. \ The Curie temperature
$T_{C}$ of the\ Gd thin film is 240 K, somewhat below the bulk value.

The absorption spectrum as a function of applied magnetic field for the
Co(1)/Ag/Co(2)/Gd multilayer is shown in Fig. 2 at room temperature. The
microwave frequency is 10 GHz and the applied field is in the plane of the
sample. Two Lorentzian derivative fits are also shown in Fig. 2 to identify
two different resonances. Separation of the adjacent absorption peaks can be
made because, as shown previously,\cite{demirtas} a proximate layer of Gd
reduces the field for resonance and significantly increases the linewidth of
Co thin films. This leads to the conclusion that the broader reasonance is
associated with the Co(2) layer. \ Fig 3 shows the temperature dependence of
the linewidth associated with Co(1) and Co(2) resonances. \ Above the Curie
temperature of Gd ($T_{C}=240$ K), both resonance lines broaden slightly with
decreasing temperature. \ Below $T_{C}$ the Co(2) resonance is no longer seen
while the Co(1) resonance first broadens abruptly and then continues to
increase with decreasing temperature to the compensation point, $T_{comp}$
$=170$ K. \ Below \ $T_{comp}$, the linewidth increases much more strongly
with decreasing temperature, exceeding the resonant field below 100\ K.%
\begin{figure}
[ptb]
\begin{center}
\includegraphics[
height=2.9032in,
width=3.5665in
]%
{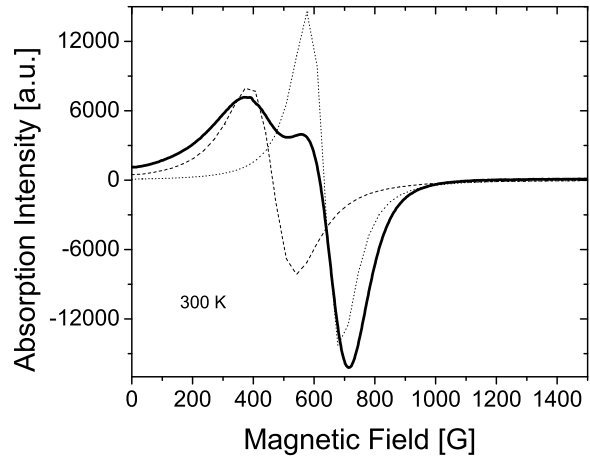}%
\caption{FMR absorption spectra for the [Co 40 \AA /Ag 100 \AA \ /Co 40
\AA /Gd 100 \AA ] film at room temperature. Linewidths were found making two
Lorentzian fits to the overall absorption spectra}%
\end{center}
\end{figure}

Ferromagnetic resonance is generally treated phenomenologically via the
Landau-Lifshitz-Gilbert (LLG) equation of motion,\cite{gilbert}
\begin{equation}
\frac{d\overrightarrow{m}}{dt}=-\gamma\overrightarrow{m}\times\overrightarrow
{H}+\alpha\overrightarrow{m}\times\frac{d\overrightarrow{m}}{dt}. \label{Eq1}%
\end{equation}
where $\overrightarrow{m}$ is the reduced magnetization vector, $\gamma,$ the
gyromagnetic ratio and $\alpha,$ the Gilbert damping parameter. \ Relaxation
in metallic ferromagnet films has conventionally been attributed to the
transfer of angular momentum from the precessing magnetization to the spin of
the conduction electrons via \textit{s-d} exchange and the subsequent
relaxation of the conduction electron polarization via spin-dependent
scattering.\cite{tserkovnyak} \ More recently, attention has been focused on
the so-called torque-correlation model first introduced by
Kambersky.\cite{kambersky} \ In this process, the time-dependent magnetization
induces charge-currents in the conduction electrons via the spin-orbit
interaction. \ These, in turn, exert torque on the magnetization, transferring
angular momentum to the lattice via the relaxation of charge currents. \ The
longer the relaxation time $\tau$ of these currents, the greater the torque
and the broader the line. \ For intraband transitions, Gilmore, et
al.\cite{gilmore08} have shown that
\begin{equation}
\alpha(T)=\frac{\gamma\tau(T)}{2\mu_{0}m}\sum_{nk}|\Gamma_{n}(k)|^{2}\left(
-\frac{\partial f}{\partial\varepsilon}\right)  , \label{Eq2}%
\end{equation}
where $\tau$ is the orbital relaxation time of the conduction electron,
$\Gamma_{n}(k)$ is the torque matrix element from the spin-orbit interaction,
and $(-\partial f/\partial\varepsilon)$ is the negative derivative of the
Fermi function. \ \ The sum is over band indices. The interplay between the
two mechanisms has been discussed by several
authors.\cite{gilmore07,tserkovnyak04}\ By artificially changing the Fermi
energy in their band calculations, Gilmore et al. demonstrate specfically that
the summation in Eq. (\ref{Eq2}) follows the density of states for Co and
other ferromagnetic metals. Note that the linewidth is related to the Gilbert
parameter by $\Delta H=1.16\omega\alpha/\gamma$, where $\omega/2\pi=10$ GHz is
the applied microwave frequency.%
\begin{figure}
[ptb]
\begin{center}
\includegraphics[
height=2.9118in,
width=3.5708in
]%
{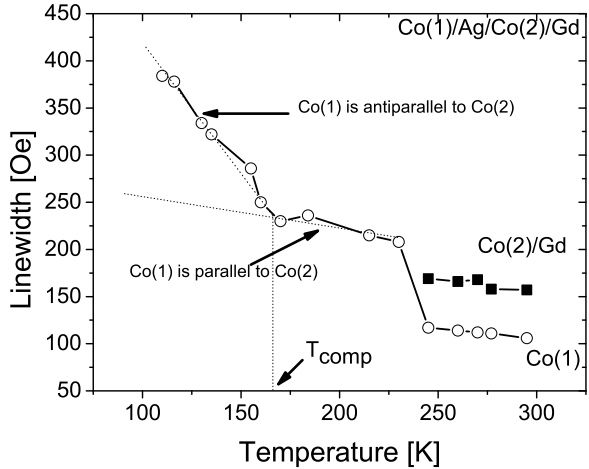}%
\caption{FMR linewidth as a function of temperature for parallel and
antiparallel alignment of Co layers in [Co 40 \AA /Ag 100 \AA \ /Co 40 \AA /Gd
100 \AA ] film. }%
\end{center}
\end{figure}

As seen in Fig. 3, the Co(1) linewidth gradually increases with decreasing
temperature from $T_{C}$ to $T_{comp}$ and then increases more rapidly below;
this is the GMB effect. \ A linewidth that increases with decreasing
temperature is indicative \cite{gilmore07} that the torque-correlation process
dominates over spin damping, evidently becoming even more dominant below
$T_{comp}$. \ In the absence of torque-correlation processes, spin-damping,
which varies $\tau(T)^{-1}$, would require a mechanism that, upon reversal of
the Co(2) magnetization, increases with decreasing temperature at a rate that
overcomes the increase in $\tau(T)$. The band structure of the Co(1) layer, on
the other hand, will change dramatically upon the transition from P to AP
alignment.\cite{schep}, thereby changing the density of states in the Co(1)
layer. Further, Binder et al.\cite{binder01} showed that impurities located
within a Co layer in GMR structures exhibit dramatically larger relaxation
rates in AP vs P alignment, again reflecting an increase in the local density
of states. Impurities located at the interface between Co and Cu, in their
calculation, are seven times more effective as scatterers in the AP
configuration; the effect is even larger for impurities in the center of the
Co layer. Similarly, the torque matrix element $\Gamma_{n}(k)$, which tracks
with the density of states, \cite{gilmore08} should reflect the same increase
in local density of states in the AP configuration. \ We attribute the
seven-fold increase in the slope of $\Delta H(T)$ shown in Fig. 3, therefore,
to an increase in the summation in Eq.(\ref{Eq2}) and consequently, to a
stronger dependence on $\tau(T)$. \ Further, Steiauf and
F\"{a}hnle\cite{steiauf} have shown, in the context of the torque-correlation
approach, that band-structure effects in lower-dimensional structures
dramatically increase the Gilbert parameter of Co relative to the bulk metal.
We suggest that, in the single layer considered by Steiauf and F\"{a}hnle,
both spin sub-bands are localized, much as in the case of AP alignment, while
only one sub-band is localized in the P configuration. \ We conclude that the
large enchancement of the temperature dependence of the linewidth in our GMR
structure--the GMB\ effect--confirms both the dominance of the
torque-correlation process in spin damping and the importance of electron
localization in the GMR effect.

\ There have been, of course, many studies of magnetic relaxation in thin
metallic films and multilayers. For example, an experiment by Urban, et
al.\cite{urban} found that\ the relaxation rate for a thin Fe layer was larger
when a second Fe layer, separated by an Au spacer, was added. Because the
increase depends on the thickness $d$ of the resonating layer, they ascribed
it to torques that occur at single ferromagnetic-normal metal interfaces, with
no role proposed for the thicker Fe layer beyond acting as a sink for spin
currents. We suggest that localization effects may play a role, even though
the conduction electrons in Fe are less polarized than in Co. A very similar
experiment \cite{heinrich} showed that when the resonance of the two layers in
an Fe/Au/Fe\ are made to coincide by judicious choice of in-plane field angle,
the linewidths are equal and narrowest. \ This was interpreted in terms of
spin pumping between the two layers. \ In that picture, the off-resonance
ferromagnetic layer acts as a perfect spin sink, except when the two layers
have a common resonant field. \ Then spin currents generated in each layer
compensate the spin-sink effect of the other.However, the resonances coincide
when the effective field is the same in each layer, which may also maximize
ferromagnetic alignment and minimize localization. As seen in Fig. 2 in the
present experiment, the Co(2) and Co(1) resonances overlap at room
temperature, and therefore each may be narrowed by spin pumping. \ Below
$T_{c},$ however, the Co(2) resonance is no longer detected, with the
antiferromagnetic coupling to the ferromagnetic moment of Gd shifting the
resonance out of the observed field range. As a consequence, we expect
dynamical coupling due to spin-pumping to disappear below the Gd transition,
giving rise to the observed jump in the linewidth of the Co(1) resonance.

To summarize, we argue that the change in the temperature dependence of the
ferromagnetic linewidth that occurs at the transition between P and AP
alignment, provides independent confirmation of the role of quantum
confinement in GMR structures. \ At the same time, it provides further
evidence that the torque-correlation model plays a substantial role in spin
relaxation in metallic ferromagnets, especially in Co, which is nearly a
half-metal. \ Clearly, similar experiments using Fe and permalloy, where the
torque-correlation model may be less dominant, are clearly in order.

One of us (ARK) wishes to acknowlege the support of the Welch Foundation
through Grant No. Y-1215.

\end{document}